\documentclass{article}
\usepackage{amsmath,graphicx,hyperref}
\usepackage[preprint]{spconf}
\usepackage{xcolor}
\usepackage{amssymb}
\usepackage[table]{xcolor}
\usepackage{booktabs}
\usepackage{multirow}
\usepackage{bm}
\usepackage{lipsum}
\usepackage{booktabs}
\usepackage{threeparttable}

\copyrightnotice{\begin{minipage}{\textwidth}
\scriptsize\copyright 2026 IEEE. Personal use of this material is permitted. Permission from IEEE must be obtained for all other uses, in any current or future media, including reprinting/republishing this material for advertising or promotional purposes, creating new collective works, for resale or redistribution to servers or lists, or reuse of any copyrighted component of this work in other works.
\end{minipage}}

\newcommand{\carflow}{\textsc{CAR-Flow}}

\title{MusicFlux: Flexible Conditioning for Text-Guided Music Generation and Editing}
\name{
Teysir Baoueb\textsuperscript{*}\quad
Leonardo Boulitreau de M. M. Marques\textsuperscript{*}\quad
Paraskevas Stamatiadis \quad
Gaël Richard
}

\address{LTCI, Télécom Paris, Institut Polytechnique de Paris, France}

\begin{document}
\ninept
\maketitle
\begin{abstract}
Music editing transforms existing recordings while preserving their original attributes. Recent approaches adapt pretrained text-to-music models for tasks like inpainting, event insertion, or restoration by conditioning them with context audio.
It was recently proposed for image editing to unify generation and editing by treating the source image as in-context information. However, applying this directly to music is computationally expensive due to long audio sequences. In this work, we develop MusicFlux, a text-guided music editor built on an efficient pretrained flow-matching text-to-audio model. We introduce two conditioning approaches that trade off editing performance and computational efficiency. Experiments show that both are competitive with strong baselines, with one achieving better editing quality and the other preserving the original sequence length and substantially improving computational efficiency.

\end{abstract}
\section{Introduction}
\label{sec:intro}
Music editing aims to transform an existing recording to meet a user-specified objective while preserving musical attributes outside the intended modification. In contrast to unconditional generation, an editor must therefore jointly satisfy a new condition and remain consistent with the source audio~\cite{liu2025melodyedit}. This concept comprises both global transformations, such as changing instrumentation or timbre, as well as temporally localized operations, including event insertion and removal and inpainting/outpainting~\cite{pan2026audio}. Such capabilities can facilitate interactive composition, music production, and post-production. 

Given the high cost of training music generators, editing systems often adapt pretrained text-to-music models by conditioning on the source audio~\cite{plitsis2024investigating}. AIRGen, for instance, combines heterogeneous adapters to MusicGen~\cite{copet2023simple} with masked training to enable arrangement, refinement, and inpainting~\cite{lin2024arrange}. Instruct-MusicGen~\cite{zhang2025instructmusicgenunlockingtexttomusicediting} augments MusicGen with audio- and text-fusion modules for instruction-guided stem addition, removal, and extraction. 
Diffusion Transformer (DiT)~\cite{peebles2023scalable} approaches have similarly introduced auxiliary controls: in ~\cite{hou2025editing}, Stable Audio Open~\cite{stableaudioopen1} is extended with a ControlNet branch conditioned on text and melody, while in Audio ControlNet~\cite{zhu2026audio}, temporal controls are introduced along with an editor for inserting or removing localized sound events. Collectively, these methods demonstrate that the generative knowledge of a pretrained model can be repurposed for editing, but generally require dedicated architecture or additional attention operations.

A simpler strategy has recently emerged in image editing. FLUX.1 Kontext~\cite{labs2025flux1kontextflowmatching} unifies generation and editing in a single flow-matching Transformer by concatenating the source image with the noisy target, allowing pretrained self-attention layers to jointly model preserved and generated content. This in-context formulation avoids task-specific architectures while supporting both global and local edits, making it a promising basis for extending text-to-music generators to editing.

\begin{figure}[t]
    \centering
    \hspace*{-0.35cm}
    \includegraphics[width=0.85\columnwidth]{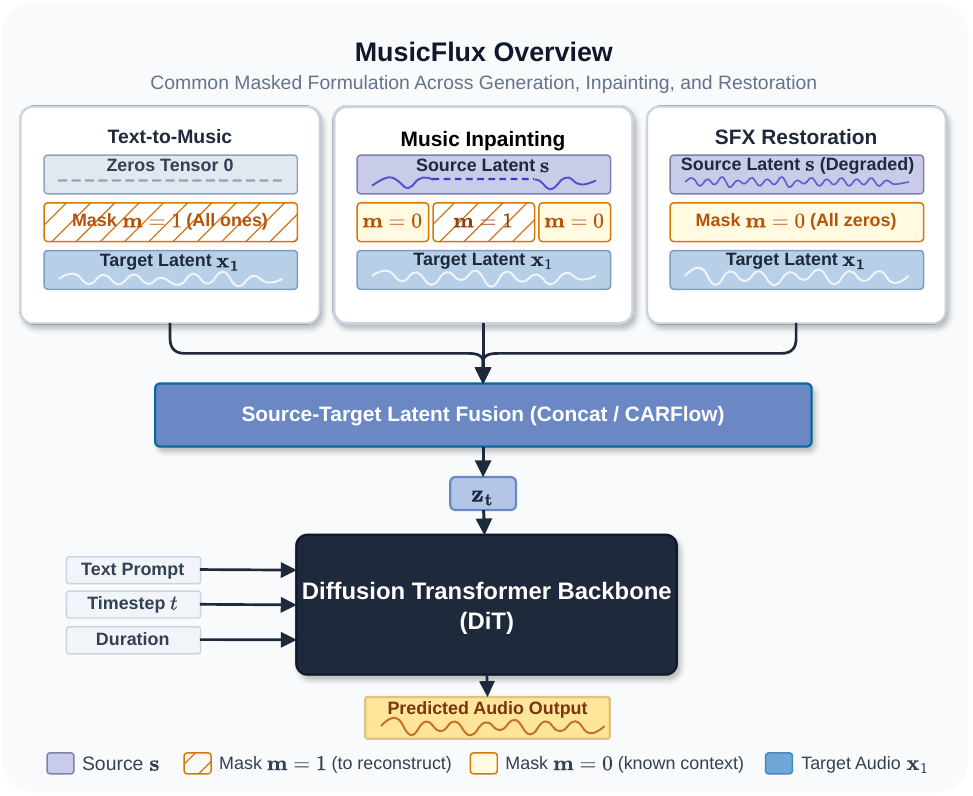}
    \caption{\textbf{MusicFlux} formulates text-to-music generation, music inpainting, and SFX restoration under a shared latent conditioning paradigm (the source latent $\textbf{s}$ represents the input audio to be transformed). The \textit{source-target latent fusion} module combines $(\textbf{s}, \textbf{m}, \textbf{x}_t)$ to construct the latent trajectory $\textbf{z}_t$ used as an input to the DiT.}
    \label{fig:overview}
\end{figure}

However, directly transferring sequence concatenation to long-form audio, thus doubling the sequence length, introduces a substantial computational burden, which reduces the feasible batch size or audio duration and compromises convergence. 
Thus, although Kontext-style concatenation provides an effective mechanism for source preservation, its direct application to music compromises the efficiency and scalability inherited from the original generator, due to its long-context nature.

In this work (see Fig.~\ref{fig:overview}), we develop a text-guided music editor based on TangoFlux~\cite{hung2026tangofluxsuperfastfaithful}, an efficient flow-matching text-to-general-audio model. We first fine-tune TangoFlux for text-to-music generation. Then, we adapt the model for music editing, in the tasks of inpainting/outpainting and sound-effects restoration by introducing source-audio conditioning without simply concatenating the source latents. Inspired by Condition-Aware Reparameterization for Flow Matching (CAR-Flow)~\cite{chen2025carflow}, we design lightweight shifting networks to predict additive shifts of the flow endpoints from the editing conditions: source audio, text prompt, and a temporal editing mask. The temporal mask explicitly identifies the regions involved in inpainting or outpainting. By embedding the editing context directly into the source and target distributions, we retain the original sequence length and relieve the flow backbone from simultaneously learning source–target transport and conditional generation.

Our contributions are threefold: 1) We specialize TangoFlux~\cite{hung2026tangofluxsuperfastfaithful} for text-to-music generation by fine-tuning its complete backbone on an open-source, large, and diverse high-fidelity dataset; 2) We extend TangoFlux into a text-guided music-editing framework by introducing joint audio, text, and temporal-mask conditioning in a straightforward approach. The resulting model supports diverse operations, including generation, localized edits such as inpainting and outpainting, and global edits such as sound effects (SFX) degradation restoration; and 3) we introduce a CAR-Flow-inspired~\cite{chen2025carflow} formulation based on condition-dependent additive transformations of the flow endpoints. In contrast to Kontext-style concatenation, CAR-Flow preserves the original sequence length, reducing attention computation and memory usage while improving efficiency. To the best of our knowledge, this is the first application of CAR-Flow to audio editing and, more broadly, to time-series data.  Our demo page is accessible at \href{https://musicflux.onrender.com/}{https://musicflux.onrender.com/}.

\section{Background}
\label{sec:background}
\subsection{Flow Matching}
Flow Matching (FM)~\cite{lipman2023flow} is a generative framework that learns a time-dependent vector field defining an ODE that continuously transports samples from a simple prior to the target data distribution. Let $\mathbf{x}_0 \sim p_0$ denote a sample from the prior and $\mathbf{x}_1 \sim p_1$ a sample from the data distribution. Given a conditional probability path $p_t(\mathbf{x}|\mathbf{x}_0,\mathbf{x}_1)$, FM learns a velocity field $\mathbf{v}_\theta(\mathbf{x},t)$ by minimizing\begin{equation}
    \mathcal{L}_{\mathrm{FM}}
    =
    \mathbb{E}_{t,\mathbf{x}_0,\mathbf{x}_1}
    \left[
        \left\|
        \mathbf{v}_\theta(\mathbf{x}_t,t)
        -
        \mathbf{u}_t(\mathbf{x}_t \mid \mathbf{x}_0,\mathbf{x}_1)
        \right\|_2^2
    \right],
\end{equation}where $\mathbf{u}_t$ is the velocity field associated with the chosen probability path. At inference, samples are generated by solving the ODE $
\mathbf{x}_t' = \mathbf{v}_\theta(\mathbf{x}_t,t)$
starting from $\mathbf{x}_0 \sim p_0$ and integrating toward $t=1$. A particularly choice is the linear interpolation between paired samples, named rectified flow, 
$\mathbf{x}_t = (1-t)\mathbf{x}_0 + t\mathbf{x}_1$,
whose conditional velocity is constant, $\mathbf{u}_t = \mathbf{x}_1-\mathbf{x}_0$. The resulting trajectories are substantially simpler than those induced by more general paths, enabling efficient generation with a small number of ODE solver steps.

\subsection{Condition-Aware Reparameterization for Flow Matching}

Standard conditional FM learns a velocity field that transports samples from a condition-independent source distribution, typically $\mathcal{N}(\mathbf{0},\mathbf{I})$, to a condition-dependent target distribution. Consequently, the velocity network must simultaneously perform probability transport and incorporate the conditioning information. CAR-Flow~\cite{chen2025carflow} addresses this by explicitly reparameterizing the source and target distributions according to the condition, thereby reducing the amount of transport that must be learned by the flow network.

In CAR-Flow, condition-dependent mappings $f(\cdot,y)$ and $g(\cdot,y)$ transform the source and target into a reparameterized space,\begin{equation}
\mathbf{z}_0 = f(\mathbf{x}_0,y),
\qquad
\mathbf{z}_1 = g(\mathbf{x}_1,y),
\end{equation}
where $y$ denotes the conditioning variable. The flow is then learned between $\mathbf{z}_0$ and $\mathbf{z}_1$, rather than directly between $\mathbf{x}_0$ and $\mathbf{x}_1$. This effectively aligns the endpoints of the probability path with the conditioning information, allowing the velocity network to focus on the remaining transport. To avoid degenerate flow solutions that can arise with unrestricted reparameterizations, CAR-Flow restricts the mappings to additive, condition-dependent shifts:
\begin{equation}
\mathbf{z}_0 = \mathbf{x}_0 + \boldsymbol{\mu}_0(y),
\qquad
\mathbf{z}_1 = \mathbf{x}_1 + \boldsymbol{\mu}_1(y),
\end{equation}
where $\boldsymbol{\mu}_0$ and $\boldsymbol{\mu}_1$ are lightweight learned networks. Thus, the condition-dependent shifts can relocate either endpoint or both. 

\subsection{TangoFlux}
TangoFlux \cite{hung2026tangofluxsuperfastfaithful} is a fast, high-fidelity text-to-audio model that generates up to 30s of 44.1,kHz audio using a flow-matching-based hybrid DiT backbone. Text is encoded with FLAN-T5 and concatenated with a learned duration embedding, while the flow timestep and pooled text representation form a global conditioning embedding that modulates the transformer through adaptive normalization. Text and duration features interact with the audio latents through joint attention in the MMDiT blocks and are subsequently processed jointly in the single-stream DiT blocks. It achieves strong performance across both objective metrics and human evaluations, while being significantly faster than many prior text-to-audio models.

\section{Proposed Method}
\label{sec:proposed_method}
We consider the problem of fine-tuning a text-to-music model to support both music generation and editing. To this end, we explore two conditioning strategies. The first, based on FLUX.1 Kontext~\cite{labs2025flux1kontextflowmatching} conditions the model by concatenating the target sequence with the contextual input. The second is inspired by CAR-Flow~\cite{chen2025carflow}, and instead introduces a condition-aware reparameterization of the model.

\subsection{MusicFlux-Concat: Context Concatenation}
\label{sec:concat}

Let $\mathbf{x}_t \in \mathbb{R}^{B\times L\times C}$ denote the noisy target latent at time $t$, where $B$ is the batch size, $L$ is the sequence length, and $C$ is the number of latent channels. Our first conditioning strategy directly extends TangoFlux by concatenating the source latent $\mathbf{s}$ with the noisy target latent along the sequence dimension (see Fig.~\ref{fig:cat_vs_car} (\textit{Left})). The source latent provides the contextual information that the model should condition on, such as an audio signal to be fully transformed or an input containing a masked region. In the absence of an audio context, as in text-to-music, the source latent is replaced by a zero tensor.

To explicitly specify how the model should use the context, we append an extra mask channel, yielding a total of $(C+1)$ channels:
\begin{equation}
\mathbf{z}_t =
\operatorname{concat}\left([\mathbf{x}_t;\mathbf{s}],[\mathbf{0};\mathbf{m}]\right)
\in \mathbb{R}^{B\times 2L\times (C+1)},
\end{equation}
where the mask $\mathbf{m}$ contains $0$ at positions corresponding to known context and $1$ at positions to be generated. On text-to-music, $\mathbf{m}$ is set entirely to $1$. For inpainting/outpainting, it contains $1$ within the region to be generated and $0$ elsewhere. For SFX restoration, where the complete audio context is provided, $\mathbf{m}$ is set entirely to $0$.

We assign distinct positional IDs to the source and target tokens, with target tokens occupying positions $0\ldots L{-}1$ and source tokens occupying positions $L\ldots2L{-}1$, following FLUX.1 Kontext~\cite{labs2025flux1kontextflowmatching}. The DiT backbone then performs full self-attention over the resulting sequence of length $2L$, while the training loss is computed only on the target half. Thus, the conditioning mechanism requires only a single additional learnable projection from $C$ to $C+1$ channels and leaves the DiT backbone architecture unchanged.

\subsection{MusicFlux-CAR: Condition-Aware Reparameterization}
\label{sec:carflow}
\begin{figure}[t]
    \centering
    \hspace*{-0.35cm}
    \includegraphics[trim=0 0 4cm 0,clip, width=1.1\columnwidth]{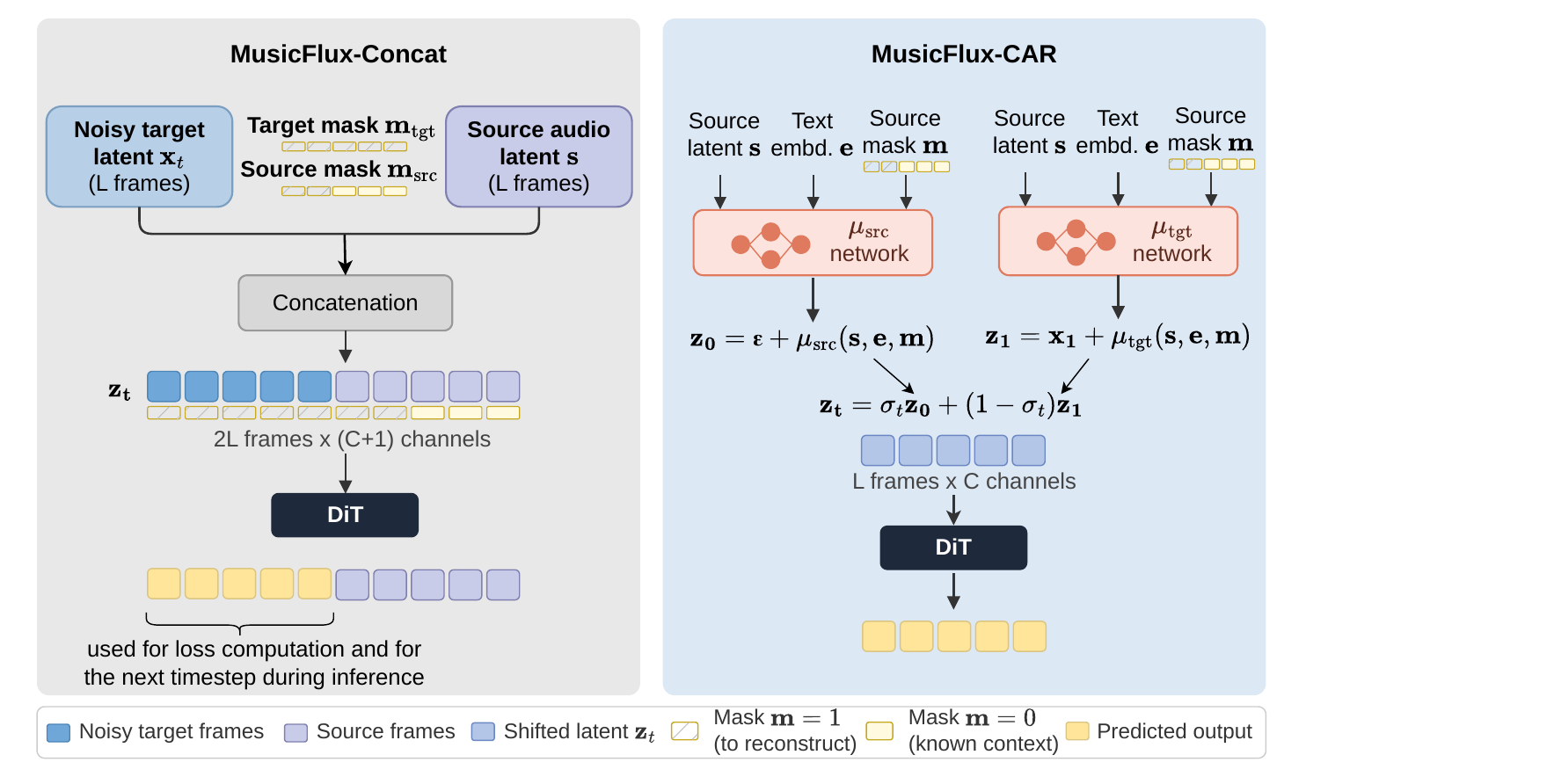}
    \caption{{\textbf{(\textit{Left})~MusicFlux-Concat:} The target latent $x_t$ and source audio latent $s$ are concatenated along the sequence dimension ($2L$ frames), 
  while the binary mask $m$ is concatenated along the channel dimension.
  \textbf{(\textit{Right})~MusicFlux-CAR:} Condition-Aware Reparameterization uses lightweight auxiliary networks ($\mu_{\text{src}}, \mu_{\text{tgt}}$) to shift the noise start-point $z_0$ and target endpoint $z_1$. Diffusion operates directly in the shifted space over $L$ frames, reducing complexity.}\vspace{-0.25cm}}
    \label{fig:cat_vs_car}
\end{figure}

The main drawback of the aforementioned approach is its computational cost since the complexity of self-attention scales with sequence length. Hence, we propose a second approach which eliminates the sequence-length doubling by pre-aligning the flow endpoints with the conditioning signal (see Fig.~\ref{fig:cat_vs_car} (\textit{Right})).
Inspired by CAR-Flow~\cite{chen2025carflow}, we introduce two lightweight \emph{Context Shift Network} modules $\mu_\text{src}$ and $\mu_\text{tgt}$ that produce additive latent shifts:\begin{equation}
  \mathbf{z}_0 = \bm{\epsilon} + \mu_\text{src}(\mathbf{s},\mathbf{e},\mathbf{m}), \hspace{1.5em}
  \mathbf{z}_1 = \mathbf{x}_1 + \mu_\text{tgt}(\mathbf{s},\mathbf{e},\mathbf{m}), 
\end{equation}where $\bm{\epsilon}\sim\mathcal{N}(\mathbf{0},\mathbf{I})$, $\mathbf{e}$ is a pooled text embedding, and $\mathbf{m}$ denotes the mask defined in the previous section.
The DiT operates on the noisy interpolant $\mathbf{z}_t=(1-\sigma_t)\mathbf{z}_1+\sigma_t\mathbf{z}_0\in\mathbb{R}^{B\times L\times C}$, which has the same length as the input used for standard unconditional text-to-music generation. 
The training loss is then given by:\begin{equation}
  \mathcal{L}_\carflow = \mathbb{E}\!\left[\|v_\theta(\mathbf{z}_t,t,\mathbf{c})-(\mathbf{z}_0-\mathbf{z}_1)\|_2^2\right].
\end{equation}At inference, the generated latent is recovered by: $\hat{\mathbf{x}}_1=\mathbf{z}_\text{final}-\mu_\text{tgt}$.
\noindent\textbf{Shift Network Architecture.}
Each shift network ($\mu_\text{src}, \mu_\text{tgt}$) processes the source context latent $\mathbf{s} \in \mathbb{R}^{B \times C \times L}$, pooled text embedding $\mathbf{e}$, and binary mask $1{-}\mathbf{m}$ using two complementary branches:\\
\textbf{(i)~Global branch}: computes the spatial mean of $\mathbf{s}$ over known-context frames, concatenates it with the text embedding $\mathbf{e}$, and passes the vector through a 2-layer MLP ($\text{Linear}\to\text{SiLU}\to\text{Linear}$) to yield a per-channel global offset broadcast across sequence length~$L$; \textbf{(ii)~Spatial branch}: processes the masked context $\mathbf{s} \odot (1{-}\mathbf{m})$ via a 1-D depthwise-separable convolution (kernel size~3) to capture local temporal details, augmented by a learnable positional bias $\mathbf{p} \in \mathbb{R}^{1 \times C \times L}$. Features from both branches are concatenated and projected via a zero-initialized linear output layer, ensuring the model begins training with identity shifts ($\mu = \mathbf{0}$) corresponding to standard unshifted flow matching endpoints. Both shift networks add only ${\approx}4.6$\,M parameters ($<0.6\%$ of the DiT backbone), maintaining negligible computational overhead.

\subsection{Task-Specific Loss Objectives}
All tasks optimize the velocity prediction $v_\theta(\mathbf{z}_t, t, \mathbf{c})$ against target flow field: $\mathbf{u}_t = \mathbf{x}_1 - \mathbf{x}_0$ for MusicFlux-Concat and $\mathbf{u}_t = \mathbf{z}_1 - \mathbf{z}_0$ for MusicFlux-CAR. For text-to-music generation and SFX restoration, we compute a full-sequence MSE loss $\mathcal{L}_{\text{full}} = \mathbb{E}\!\left[\|v_\theta - \mathbf{u}_t\|^2\right]$. For inpainting and outpainting, the objective combines two distinct terms over the target reconstruction region $\mathbf{m}$ and context region $(1-\mathbf{m})$, $ \mathcal{L}_{\text{inpaint}} = \mathcal{L}_{\text{masked}} + \lambda_{\text{ctx}} \, \mathcal{L}_{\text{ctx}},
$ where $\mathcal{L}_{\text{masked}} = \mathbb{E}\!\left[\mathbf{m} \odot \|v_\theta - \mathbf{u}_t\|^2\right]$ penalizes errors within the generated region, while $\mathcal{L}_{\text{ctx}} = \mathbb{E}\!\left[(1-\mathbf{m}) \odot \|v_\theta - \mathbf{u}_t\|^2\right]$ encourages fidelity to the context boundaries. We use a factor $\lambda_{\text{ctx}} < 1.0$ for the context loss, since reconstructing the observed context is more straightforward.   

\begin{table*}[!ht]
    \centering
    \resizebox{\textwidth}{!}{%
    \begin{tabular}{l
                    cc cc
                    cc cc
                    cccc}
        & \multicolumn{4}{c}{\textbf{Generation}}
        & \multicolumn{4}{c}{\textbf{Inpainting/Outpainting}}
        & \multicolumn{4}{c}{\textbf{Restoration}} \\
        \cmidrule(lr){2-5}
        \cmidrule(lr){6-9}
        \cmidrule(lr){10-13}

        & \multicolumn{2}{c}{JMC-Test}
        & \multicolumn{2}{c}{SDD}
        & \multicolumn{2}{c}{JMC-Test}
        & \multicolumn{2}{c}{SDD}
        & \multicolumn{4}{c}{SonicMaster-Test} \\
        \cmidrule(lr){2-3}
        \cmidrule(lr){4-5}
        \cmidrule(lr){6-7}
        \cmidrule(lr){8-9}
        \cmidrule(lr){10-13}

        \textbf{Model}
        & CAF $\uparrow$ & KAD $\downarrow$
        & CAF $\uparrow$ & KAD $\downarrow$
        & FLEUR* $\uparrow$ & KAD $\downarrow$
        & FLEUR* $\uparrow$ & KAD $\downarrow$
        & KAD $\downarrow$ & KL $\downarrow$ & SSIM $\uparrow$ & PQ $\uparrow$ \\
        \midrule

        Ground-truth (GT)
        & \underline{0.5102} & \textbf{0.0000}
        & \textbf{0.5704} & \textbf{0.0000} 
        & \textbf{0.8592} & \textbf{0.0000} 
        & \textbf{0.8834} & \textbf{0.0000} 
        & \textbf{0.0000} & \textbf{0.0000} & \textbf{1.000} & \textbf{7.7951} \\

        Stable Audio 3~\cite{evans2026stableaudio3}
        & -- & -- 
        & \underline{0.5593} & \underline{2.8631} 
        & -- & -- 
        & 0.8248 & 1.4794  
        & -- & -- & -- & -- \\

        SonicMaster~\cite{melechovsky2025sonicmastercontrollableallinonemusic}
        & -- & -- 
        & -- & -- 
        & -- & -- 
        & -- & -- 
        & 0.1137 & \underline{0.0009} & \underline{0.5918} & 7.5310 \\

        \rowcolor{gray!15}
        \textbf{MusicFlux-Concat}
         & 0.5100 & \underline{0.6400} 
        & 0.5213 & 4.5535 
        & \underline{0.8340} & \underline{0.0099} 
        & \underline{0.8428} & \underline{0.0380} 
        & \underline{0.0489} & 0.0013 & 0.5702 & \underline{7.5478} \\

        \rowcolor{gray!15}
        \textbf{MusicFlux-CAR}
        & \textbf{0.5196} & 1.1239 
        & 0.4950 & 4.6687 
        & 0.8234 & 0.2496 
        & 0.8390 & 0.4744 
        & 0.2680 & 0.0015 & 0.5412 & 7.4044 \\

        \bottomrule
    \end{tabular}%
    }
    \caption{Evaluation results of music generation, inpainting, and restoration.
    Best results are shown in bold and second best are underlined.}
    \label{tab:objective_results}
\end{table*}

\section{Experiments}
\label{sec:experiments}
\subsection{Datasets}
\textbf{JamendoMaxCaps (JMC)}~\cite{roy2025jamendomaxcapslargescalemusiccaption}: A collection of instrumental audio collected from the Jamendo Music platform paired with metadata and captions generated using Qwen2-Audio~\cite{chu2024qwen2}. The corpus contains approximately 362K tracks, totaling around 58K hours, with an average duration of 4 minutes and 12 seconds per track. All audio is stereo and standardized at 44.1 kHz. Each 30-second audio segment is captioned, producing 1,812,341 paired audio–caption samples. We split the dataset into 357 samples for validation, 10,000 for testing, and the remainder is allocated for training.\\
\textbf{SonicMaster-Test}~\cite{melechovsky2025sonicmastercontrollableallinonemusic}: 1,000 clean 30-s music excerpts and 7,000 corresponding degraded versions, with seven variants generated per reference: four single-, two double-, and one triple-degradation examples. The degradations cover 19 audio effects from five categories: equalization, dynamics, reverb, amplitude, and stereo. Natural language instructions for restoration are given.\\
\textbf{Song Describer Dataset (SDD)}~\cite{manco2023song}: 706 openly licensed music recordings paired with 1,106 human-written captions, with 95\% of the audio excerpts lasting up to 2 minutes. In our experiments, we use the \textbf{valid} subset, which went through a manual quality-control process and further exclude tracks whose captions indicate the presence of vocals, yielding an instrumental-only evaluation subset. 

\subsection{Setup}
We use the JMC training split first fine-tune the TangoFlux model for music generation, and then to train the editing models. These were trained for 4 epochs, on 3 L40S GPUs, with a per-device batch size of 32 for MusicFlux-CAR and 16 for the MusicFlux-Concat version with the Adam optimizer. We set a learning rate of $2e^{-4}$, with a 2k warmup and a cosine-with-restarts scheduler with 2 restarts. We sample all three tasks with equal probability. We use a Transformer with 6 dual-stream and 18 single-stream blocks, 8 heads of dimension 128, and a joint conditioning dimension of 1024. In total, MusicFlux-CAR and MusicFlux-Concat have roughly 859M and 860M parameters, respectively. The model operates on sequences of maximum 645 latents, extracted with Stable Audio Open 1.0~\cite{stableaudioopen1}, corresponding to 30 seconds. We found that $\lambda_{\text{ctx}}=0.1$ yielded the best results. We select the checkpoint that performs best jointly on both generation and editing, according to a CAF-Score metric~\cite{lee2026cafscorecalibratingclaplalms} on the JMC validation set.  The SonicMaster-Test and the SDD dataset are used as test sets for out-of-domain evaluation of SFX restoration and inpainting/outpainting and generation, respectively. We run inference with 50 integration steps and a CFG scale of 4.5.

\subsection{Baselines}
For music generation and inpainting/outpainting, we use the state-of-the-art Stable Audio 3.0 (SA3)~\cite{evans2026stableaudio3} as a baseline. The model is also based on a DiT encapsulated in a Flow Matching framework. Frozen T5Gemma text embeddings, concatenated with a duration embedding, condition the model through cross-attention. Instead of using CAR-Flows or concatenation, the context audio is linearly projected and added to the hidden states of each transformer block. We use its \textit{small-music-base} version, as it's the closest in terms of size to MusicFlux, with 459M parameters. It was trained on a mixture of the private, curated AudioSparx and Freesound datasets, with $\approx$1.2M tracks. For restoration, we use SonicMaster~\cite{melechovsky2025sonicmastercontrollableallinonemusic}, a text-conditioned rectified-flow model that directly transports degraded audio latents toward clean ones. Unlike MusicFlux, the degraded source is not provided as an additional conditioning signal, but instead serves as the initial state of the flow trajectory.

\begin{table}[t]
    \centering

    \begin{threeparttable}
        \begin{tabular}{llrrr}
            \toprule
            \textbf{Stage} &
            \textbf{MusicFlux} &
            \textbf{Base} &
            \textbf{Dynamic} &
            \textbf{RTF} \\
            & \textbf{Method} & \textbf{(MiB)} & \textbf{(MiB)} & \\
            \midrule
            \multirow{2}{*}{Inference}
                & CAR & 3364.19 & \textbf{75.71}  & \textbf{0.1188} \\
                & Concat   & \textbf{3358.54} & 150.11         & 0.2203          \\
            \midrule
            \multirow{2}{*}{Training}
                & CAR & 3372.48 & \textbf{1988.10} & -- \\
                & Concat   & \textbf{3366.82} & 2748.19          & -- \\
            \bottomrule
        \end{tabular}
    \end{threeparttable}
    \caption{
        Computational comparison MusicFlux versions. Measurements use a batch size of~1, a sequence of 645, and FP32 weights.
    }
    \label{tab:efficiency}
\end{table}

\subsection{Evaluation Metrics}
For generation we report \textbf{CAF-Score}~\cite{lee2026cafscorecalibratingclaplalms} for text--audio alignment and \textbf{Kernel Audio Distance (KAD)}~\cite{chung2025kad} for distributional audio quality. CAF-Score combines CLAP similarity with FLEUR-based LALM assessment~\cite{lee-etal-2024-fleur}; we use Speech-Music LAION CLAP. KAD computes the MMD between CLAP-LAION music embeddings of generated and pre-editing audio. For inpainting/outpainting, the objective is seamless continuation of the source rather than caption alignment. We therefore replace CAF-Score with \textbf{FLEUR*}, a task-specific modification of FLEUR~\cite{lee-etal-2024-fleur}. Given the boundaries of the regenerated region, Qwen3-Omni-30B-A3B-Instruct \cite{xu2025qwen3omni} rates from 0 to 1 how well it fits the recording, considering boundary continuity, timbre, loudness, ambience, rhythm, and audible artifacts. We query three settings: (i) edited audio alone, (ii) edited audio paired with the masked source, and (iii) edited audio paired with the unmasked ground truth. Following FLEUR, ratings are smoothed using score-token probabilities, and FLEUR* is their average. We additionally report KAD for inpainting/outpainting. For restoration, we use the perceptual fidelity and production-quality metrics adopted by SonicMaster~\cite{melechovsky2025sonicmastercontrollableallinonemusic}. \textbf{KL divergence} measures distributional discrepancy, \textbf{SSIM} measures structural similarity between 128-bin mel-spectrograms, and \textbf{Production Quality (PQ)} from Audiobox Aesthetics~\cite{tjandra2025metaaudioboxaestheticsunified} estimates perceived technical quality in aspects such as clarity, dynamics, balance, and spatialization.

\section{Results}
\label{sec:results}
Table~\ref{tab:objective_results} summarizes the results for all metrics used. The first row reports the results on the ground-truth (GT) test set, where each metric is computed by comparing the ground-truth audio or set against itself. For in-domain generation on JMC-Test, MusicFlux-CAR achieves the best CAF-Score (0.5196), just above MusicFlux-Concat and GT, while MusicFlux-Concat obtains the lowest KAD (0.6400). This shows that both conditioning strategies preserve the generation capabilities of the pretrained model while adapting to new tasks, with CAR favoring text-audio alignment and Concat distributional similarity. On the out-of-domain SDD, SA3 obtains better results in both metrics, showing that the proposed models generalize less to unseen prompts. This could be due to better prompt extraction, given the curated dataset of SA3, or better prompt encoding.

For inpainting/outpainting, MusicFlux-Concat provides the strongest results in both metrics on all in- and out-of-domain scenarios. MusicFlux-CAR also beats the SA3 baseline on the out-of-domain evaluation. This advantage of MusicFlux-Concat over all other models might come from the direct exposure of the context to the attention layers, whereas in MusicFlux-CAR the context is conveyed through lightweight additive shifts, and in SA3 through a simple projection and sum, both of which could potentially be acting as information bottlenecks for localized reconstruction.

For SFX restoration, MusicFlux-Concat achieves the best KAD and PQ, while SonicMaster performs best in KL divergence and SSIM. Both MusicFlux variants remain competitive with the specialized SonicMaster baseline despite being trained jointly across multiple tasks, with MusicFlux-Concat slightly outperforming MusicFlux-CAR. This advantage may again result from the direct access to source information provided by sequence concatenation, whereas CAR encodes the context indirectly through latent shifts, making fine-grained restoration more challenging.

Lastly, Table~\ref{tab:efficiency} shows that CAR substantially improves computational efficiency over Concat. RTF and Dynamic memory (model execution) are reduced by over 50\% at inference and 28\% at training, while base memory (model instantiation) is kept almost the same due to the lightweight shift networks. This confirms the benefit of preserving the original sequence.

\section{Conclusion}
\label{sec:conclusion}
We introduced MusicFlux, a unified framework covering three music editing settings: global text-to-music generation, localized editing through inpainting/outpainting, and global editing through SFX restoration. We explored two conditioning strategies: MusicFlux-Concat, which achieves stronger editing performance, and MusicFlux-CAR, which preserves the original sequence length and reduces inference runtime and dynamic memory by over 50\%. Both remain competitive with task-specific baselines despite being single multi-task models trained on open data. The results highlight a trade-off between editing fidelity and efficiency, while limitations in out-of-domain generalization and CAR conditioning motivate richer efficient conditioning in future work.

\section{Acknowledgements}
With support from the European Union (ERC, HI-AUDIO, \linebreak 101052978). Views and opinions expressed are, however, those of the authors only and do not necessarily reflect those of the European Union or the European Research Council. Neither the European Union nor the granting authority can be held responsible for them. This work was performed using HPC resources from GENCI-IDRIS (Grants 2026-AD011018031, 2026-AD011017871, 2026-AD011014567R2, and 2026-AD011014567R3) and was supported by the IDRIS Open Hackathon 2026 program in collaboration with NVIDIA and OpenACC-Standard.org. The authors thank Martin Comminges, Krishna Kalyan and Némo Bouillon for their contributions.
\vspace{-1em}

%\vfill
%\pagebreak

% \vfill

% \pagebreak
% \clearpage

\bibliographystyle{IEEEbib}
\bibliography{refs}

\end{document}